\documentclass[]{aastex631}

\shorttitle{Electron/proton energization in 3D reconnection}
\shortauthors{Werner \& Uzdensky}

\usepackage{xpatch}
\xpatchcmd{\thebibliography}{\twocolumngrid}{}{}{}

\begin{document}

\title{Electron and proton energization in 3D reconnecting current sheets in semirelativistic plasma with guide magnetic field
}

\correspondingauthor{Gregory R. Werner}
\email{Greg.Werner@colorado.edu}

\author[0000-0001-9039-9032]{Gregory R. Werner}
\affiliation{
Center for Integrated Plasma Studies, Physics Department, \\
390 UCB, University of Colorado, Boulder, CO 80309, USA
}

\author[0000-0001-8792-6698]{Dmitri A. Uzdensky}
\affiliation{
Center for Integrated Plasma Studies, Physics Department, \\
390 UCB, University of Colorado, Boulder, CO 80309, USA
}

%
%
%
%
%



\begin{abstract}

Using 3D particle-in-cell simulation, we characterize energy conversion, as a function of guide magnetic field, in a thin current sheet in semirelativistic plasma, with relativistic electrons and subrelativistic protons. There, magnetic reconnection, the drift-kink instability (DKI), and the flux-rope kink instability all compete and interact in their nonlinear stages to convert magnetic energy to plasma energy. We compare fully 3D simulations with 2D in two different planes to isolate reconnection and DKI effects. In zero guide field, these processes yield distinct energy conversion signatures: ions gain more energy than electrons in 2D$xy$ (reconnection), while the opposite is true in 2D$yz$ (DKI), and the 3D result falls in between. The flux-rope instability, which occurs only in~3D, allows more magnetic energy to be released than in~2D, but the rate of energy conversion in 3D tends to be lower. Increasing the guide magnetic field strongly suppresses DKI, and in all cases slows and reduces the overall amount of energy conversion; it also favors electron energization through a process by which energy is first stored in the motional electric field of flux ropes before energizing particles. Understanding the evolution of the energy partition thus provides insight into the role of various plasma processes, and is important for modeling radiation from astrophysical sources such as accreting black holes and their jets.


\end{abstract}



\section{Introduction} \label{sec:intro}

In many plasma environments, magnetic energy is converted to plasma energy at thin current sheets (CSs), likely via magnetic reconnection \citep[e.g.,][]{Zweibel_Yamada-2009}.
Reconnection occurs in diverse regimes and, even in 2D, generates such complexity that we owe much of our understanding to numerical simulation.
Unfortunately, 3D simulation studies are far costlier than~2D, motivating efforts to understand similarities between 2D and 3D reconnection and more generally to understand 3D CS evolution in terms of 2D processes whenever possible.

A thin CS, considered in 2D [specifically in ``2D$xy$'', the $x$-$y$ plane perpendicular to the current~$j_z$ supporting a magnetic field $\hat{\bf x}B_0 \textrm{tanh}(y/\delta)$ reversing over thickness~$2\delta$], will typically be most unstable to tearing, leading, in its nonlinear stage, to reconnection and rapid magnetic energy release.
In 2D$yz$, the same CS may kink or ripple due to the drift-kink instability (DKI), which in its nonlinear stage induces severe contortions that rapidly release magnetic energy 
\citep[e.g.,][]{Pritchett_etal-1996}.  The DKI requires two counter-drifting species, hence is not a magnetohydrodynamics (MHD) instability; it is important when the separation between electron and ion scales is small \citep{Daughton-1999,Hesse_Birn-2000,Scholer_etal-2003}, or absent as in relativistic pair plasma \citep{Zenitani_Hoshino-2005b,Zenitani_Hoshino-2008,Yin_etal-2008,Liu_etal-2011,Cerutti_etal-2014b,Guo_etal-2014,Guo_etal-2015}.
In 3D, these instabilities can compete and interact. 
Although 2D-like reconnection may dominate DKI in~3D \citep{Kagan_etal-2013,Sironi_Spitkovsky-2014,Guo_etal-2014,Guo_etal-2015,Cerutti_etal-2014b,Werner_Uzdensky-2017,Guo_etal-2021},
\citet{Werner_Uzdensky-2021} highlighted how the DKI can dramatically slow reconnection.
Furthermore, in 3D only, flux ropes formed by reconnection can decay via the MHD kink instability \citep{Kagan_etal-2013,Markidis_etal-2014,ZhangQ_etal-2021,Schoeffler_etal-2023},
releasing additional magnetic energy \citep{Werner_Uzdensky-2021}.

Using particle-in-cell (PIC) simulation, this work characterizes energy conversion as a function of guide magnetic field, in 3D CSs in magnetically-dominated, collisionless electron-proton plasma, with true mass ratio $\mu=1836$, in the semirelativistic regime.  In semirelativistic plasma, where electrons are ultrarelativistic and protons subrelativistic, relativistic effects diminish the scale separation between (e.g., gyroradii of) electrons and ions (protons), making DKI consequential.
The semirelativistic regime is one of the several important regimes for black hole (BH) accretion flows, which are expected to have electron-ion plasma at temperatures around~10--100$\:$MeV \citep[e.g.,][]{Dexter_etal-2020a} and likely develop large, complicated, turbulent magnetic field structures in the accretion flow, in the overlying corona/wind, and in and around the jet \citep[e.g.,][]{Galeev_etal-1979,Uzdensky_Goodman-2008,Ripperda_etal-2020,Chashkina_etal-2021,Bacchini_etal-2022,Zhdankin_etal-2023}.  We generally expect CS formation and reconnection in such contexts, and the resulting particle energization could potentially power flaring emission in BHs (as in solar flares).
Furthermore, characterizing electron and ion energization in the semirelativistic regime is crucial for global MHD modeling of reconnection-powered emission from accreting BHs 
\citep{Chael_etal-2018,Dexter_etal-2020a,Ressler_etal-2020,Ressler_etal-2023,Scepi_etal-2022,Hankla_etal-2022}.

Until now (to our knowledge), no 3D PIC studies of electron-ion reconnection (or CS evolution) have characterized energy conversion in the nonradiative semirelativistic regime \citep[N.B.,][includes 1\% ultrarelativistic ions among positrons and electrons, with radiative cooling]{Chernoglazov_etal-2023}, although 
several 2D reconnection studies have \citep{Melzani_etal-2014a,Melzani_etal-2014b,Guo_etal-2016,Rowan_etal-2017,Rowan_etal-2019,Werner_etal-2018,Ball_etal-2018,Ball_etal-2019}.

After describing the simulation setup (\S\ref{sec:setup}), we will focus on the different energy components, discussing the release of magnetic energy~$U_{Bxy}$ (\S\ref{sec:dUBxy}),  the increase in guide field energy~$U_{Bz}$ (\S\ref{sec:dUBz}), the electric field energy~$U_E$ (\S\ref{sec:dUE}), and the electron and ion energies~$U_e+U_i=U_{\rm ptcl}$ (\S\ref{sec:dUptcl}).
We summarize in~\S\ref{sec:summary}.

\section{The simulations} \label{sec:setup}

Simulations were run with the {\sc Zeltron} electromagnetic PIC code \citep{Cerutti_etal-2013},
initialized with semirelativistic electron-proton plasma in a standard Harris sheet configuration with uniform background plasma \citep[similar to][]{Werner_etal-2018}---initial physical and numerical parameters are listed in Table~\ref{tab:initPlasma}. By ``semirelativistic'' we mean that ions remain subrelativistic while electrons are ultrarelativistic and experience ultrarelativistic energy gains; in terms of quantities in Table~\ref{tab:initPlasma}, $m_e c^2 \ll T_b \ll m_i c^2$ and $\sigma_i \ll 1 \ll \sigma_e$. The simulation frame (with zero initial electric field) is the zero-momentum frame, where ions drift slowly and electrons carry the current supporting the field reversal.

\begin{table}[h]
  \caption{\label{tab:initPlasma}
    Initial plasma and numerical simulation parameters.
  }
\begin{tabular}{ll}
  \hline
  $ \mu \equiv m_i/m_e = 1836 $ & ion/electron mass ratio \\
  $B_0$ & asymptotic upstream (reversing) magnetic field \\
  $n_b=n_{be}=n_{bi}$ & 
    background electron/ion density \\
  $T_{b} = 18.36 m_e c^2 = 0.01 m_i c^2 $
    &
    background electron/ion temperature \\
  $\sigma_i \equiv B_0^2 / (4\pi n_b m_i c^2) = 0.5$ & 
    background ion cold magnetization\\
  $\sigma_e \equiv B_0^2 / (4\pi n_b m_e c^2) = \mu \sigma_i=918$ & 
    background electron cold magnetization\\
  $ \delta=0.33 \sigma_i \rho_{i0}$ & 
    Harris CS half thickness ($\rho_{i0}\equiv m_i c^2/eB_0$)\\
  $\mathbf{B}(y) = \hat{\bf z}B_{gz} - 
                \hat{\bf x} B_0 \textrm{tanh}(y/\delta)$ & 
    magnetic field profile \\
  $-0.6 c\hat{\bf z}$, $+0.0004 c \hat{\bf z}$  & 
    electron, ion bulk drift velocities in CS \\
  $\eta\equiv n_{d,\rm max}/n_b=5$ & CS overdensity \\
  $n_{d}(y)= \eta n_b \textrm{cosh}^{-2}(y/\delta)$ &
    CS electron/ion density profile \\
  $T_d = 51 m_e c^2 = 0.28 m_i c^2$ & 
    CS electron/ion (co-moving) temperature  \\
  $\beta_{be} = \beta_{bi} \equiv 8\pi n_b T_b / B_0^2 = 0.04$ &
    upstream plasma beta for electrons/ions \\
  $v_A = 0.57c$ &
    upstream Alfv\'{e}n velocity for $B_{gz}=0$\\
  $L_y \equiv L = 55.3\sigma_i \rho_{i0}$ & system size (in $y$)  \\
  $L_x/L = 1$ & system $x/y$ aspect ratio (for 2D$xy$ and 3D) \\
  $L_z/L = \textrm{max}(1, B_{gz}/B_0) $ & 
    system $z/y$ aspect ratio (for 2D$yz$ and 3D)  \\
  $ T=20L/c$ & simulation duration \\
  $\Delta y=\Delta z = 0.036 \sigma_i \rho_{i0}$ & grid cell size in $y$ and $z$ \\
  $\Delta x= 0.058 \sigma_i \rho_{i0}$ & grid cell size in $x$ \\
  $\Delta t = 0.64\Delta y/c$ & 
    timestep (the 3D Courant-Friedrichs-Lewy maximum) \\
  $M_{\rm ppcs}=10$ (3D), 160 (2D) & macroparticles per cell per species (4 species) \\
  $M_{\rm ppc\phantom{s}}=40$ (3D), 640 (2D) & total macroparticles per cell \\
  \hline
\end{tabular}
\end{table}

\begin{table}[h]
  \caption{\label{tab:lengthScales}
    Initial length scales normalized to 
      $\sigma_i \rho_{i0}=\sigma_e \rho_{e0}$, where 
      $\rho_{s0}=m_s c^2/eB_0$, $\sigma_s=B_0^2/(4\pi n_{bs} m_s c^2)$
    }
  \begin{tabular}{lrl}
    \hline
    $\Delta y=\Delta z$ & 0.036 & grid cell size in $y$, $z$ \\
    $\Delta x$ & 0.058 & grid cell size in $x$ \\
    $\rho_{be}$ & 0.060   & upstream electron gyroradius 
      ($B_{gz}=0$) \\
    $\lambda_{De}=\lambda_{Di}$ & 0.14\phantom{6}   & 
      upstream electron/ion Debye length \\
    $d_e$ & 0.25\phantom{6}   & upstream relativistic electron skin depth \\
    $\delta$ & 0.33\phantom{6}   & CS half-thickness \\
    $\rho_{bi}$ & 0.35\phantom{6}  & upstream ion gyroradius 
      ($B_{gz}=0$) \\
    $\sigma_e \rho_{e0}$ & 1.0\phantom{06} &
      twice the energized electron gyroradius, 
      if all magnetic energy went to electrons \\
    $d_i$ & 1.4\phantom{06} & upstream nonrelativistic ion skin depth \\
    $L$ & 55.3\phantom{06} & system size in $x$, $y$, and $z$ 
           (except: if $B_{gz}/B_0>1$,
           then $L_z/L=B_{gz}/B_0$) \\
    \hline
  \end{tabular}
\end{table}

The initial CS is perpendicular to $y$, with current in the $+z$ direction and reversing magnetic field in the~$\pm x$ directions (approaching value $B_0$ upstream).
We add uniform guide field $B_{gz}\hat{\bf z}$,
exploring several strengths $B_{gz}/B_0 \in \{0,0.25,0.5,1,2,4\}$ (except $B_{gz}/B_0=4$ was not run in~3D, due to expense).  We also explore three dimensionalities: 2D$xy$ (2D simulation ignoring~$z$---classic 2D reconnection), 2D$yz$ (ignoring~$x$, for studying DKI), and full~3D.
Also: in one case, 2D$yz$ and $B_{gz}=0$, we ran an ensemble of 3 simulations, identical except for differing randomization of initial particles.

Table~\ref{tab:lengthScales} lists length scales of interest in terms of $\sigma_i \rho_{i0}=\sigma_e \rho_{e0}$, where $\rho_{s0}\equiv m_s c^2/(eB_0)$; $\sigma_e \rho_{e0}$ is the gyroradius of an ``energized'' electron with energy $2 B_0^2/(8\pi n_{be})$ in perpendicular field~$B_0$.

The system size, containing a single CS, is $L \equiv L_y=55.3\sigma_i \rho_{i0}$ and $L_x = L_z = L$ [except when $B_{gz}>B_0$, we increase $L_z=(B_{gz}/B_0)L_y$; and 2D simulations lack either
the $x$ or $z$ dimension]. Boundary conditions are periodic in $x$ and
$z$, and conducting/particle-reflecting in~$y$.
The size~$L$ is large with respect to initial kinetic scales and CS thickness [e.g., $L/(2\delta) = 83$], and $L > 20 \sigma_i \rho_{i0}$ puts it in the ``large system regime'' as defined by \citet{Werner_etal-2016}, albeit for relativistic pair plasma in 2D$xy$.

We empirically found that using anisotropic grid cells with
$\Delta x/1.6 = \Delta y=\Delta z = \sigma_i \rho_{i0}/28$
avoids numerical instabilities while yielding the best computational performance.
We similarly found that the number of computational macroparticles (each representing many physical particles) needed to avoid unphysical electron-ion energy exchange at late times was 
160 per cell per species (with 4 species: background and CS electrons and ions) in~2D, whereas, in~3D, 10 macroparticles/cell/species sufficed.

No initial perturbation is used to kickstart reconnection, because such a perturbation (if uniform in~$z$) can significantly suppress 3D effects \citep{Werner_Uzdensky-2021}.  Initial instabilities are therefore seeded by macroparticle noise.

\section{Qualitative evolution and transverse magnetic energy released} \label{sec:dUBxy}

\begin{figure}
\centering
\raisebox{0.1in}{(a)}%
\includegraphics*[width=3in]{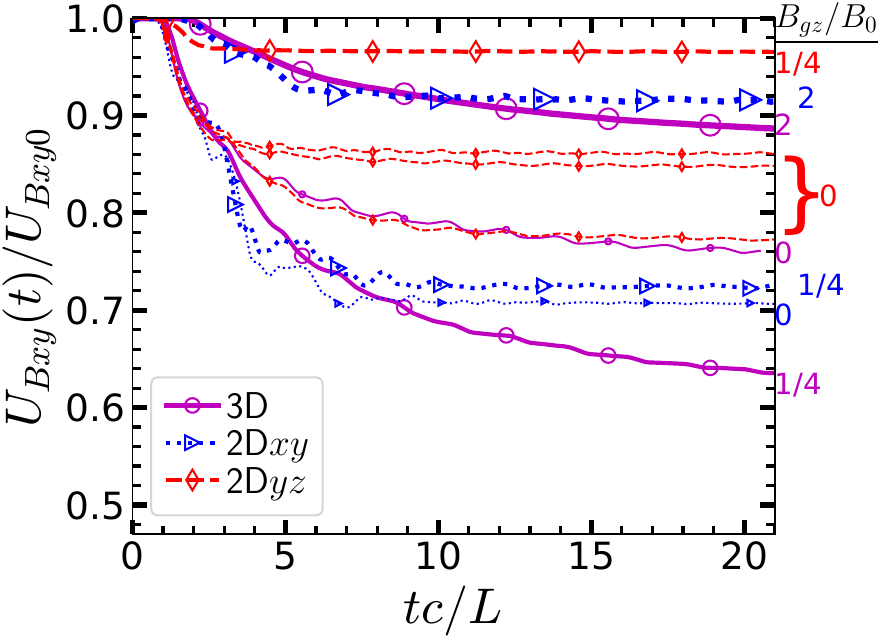}
\raisebox{0.1in}{(b)}%
\includegraphics*[width=3in]{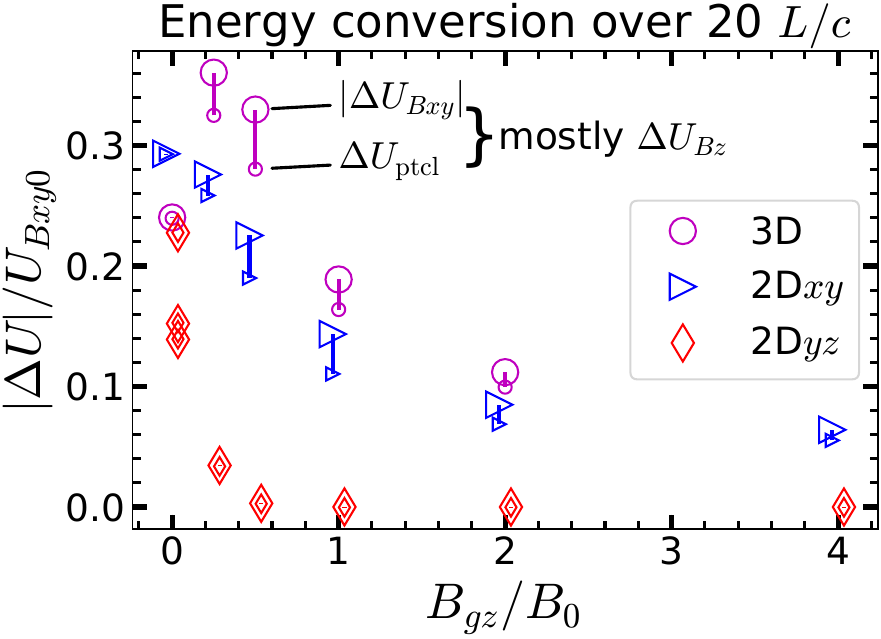}
\caption{\label{fig:energyConversion}
(a) $U_{Bxy}(t)$ for selected simulations
($B_{gz}$ is labeled for each).
Three (differently-seeded) 2D$yz$ simulations with $B_{gz}=0$ demonstrate the enhanced stochastic variability of DKI.
(b)
$|\Delta U_{Bxy}|$ (large symbols) and $\Delta U_{\rm ptcl}$ (small symbols) versus $B_{gz}$, 
normalized to $U_{Bxy0}$, the initial transverse magnetic energy.
}
\end{figure}

Over time, transverse magnetic energy~$U_{Bxy}$ is released to other forms ($U_{Bxy}$ excludes energy in guide field~$B_z$, which cannot be released---see~\S\ref{sec:dUBz}).
Figure~\ref{fig:energyConversion}(a) shows the evolution of $U_{Bxy}(t)$ for selected simulations.
Figure~\ref{fig:energyConversion}(b) shows
the net loss in $U_{Bxy}$ over $T=20L/c$ as a function of~$B_{gz}$,
for all simulations (large symbols).
Ultimately this energy~$\Delta U_{Bxy}$ goes almost entirely to particles ($\Delta U_{\rm ptcl}$, small symbols; cf.~\S\ref{sec:dUptcl}), with the small
difference $|\Delta U_{Bxy}| - \Delta U_{\rm ptcl}$ comprising mainly $\Delta U_{Bz}$, the increase in guide-field energy (see~\S\ref{sec:dUBz}).

In 2D$xy$ (Fig.~\ref{fig:energyConversion}a, blue triangles), magnetic energy decreases rapidly due to magnetic reconnection until saturation after 5--10$\:L/c$.  During that time we observe the familiar behavior of plasmoid-dominated reconnection.
Reconnection pumps magnetic energy and flux from the upstream into plasmoids (magnetic islands), continually growing the total plasmoid volume and the magnetic energy therein.  Apart from merging, plasmoids are permanent (stable) in~2D.  Eventually (after 5--10$\:L/c$), in a closed system, the largest plasmoid becomes so large ($\sim\!\!L$) that it halts reconnection---even though considerable transverse magnetic energy may remain upstream as well as in the plasmoid.

In 2D$yz$ (Fig.~\ref{fig:energyConversion}, red diamonds), tearing and reconnection are forbidden, but the nonlinear DKI can rapidly release magnetic energy.  We observe qualitatively similar behavior---the rippling of the CS, sometimes leading to a catastrophic folding-over on itself---as reported in nonrelativistic electron-ion plasma with artificially low $m_i/m_e$ and in relativistic pair plasma \citep{Ozaki_etal-1996,Pritchett_Coroniti-1996,Pritchett_etal-1996,Zhu_Winglee-1996,Zenitani_Hoshino-2007,Cerutti_etal-2014b,Werner_Uzdensky-2021}.
The nonlinear DKI effectively thickens the CS, significantly slowing but not fully halting energy conversion.
A DKI mode may enter the nonlinear stage when its rippling amplitude exceeds its rippling wavelength, but---like waves breaking in the ocean---whether and where this occurs has a substantial element of randomness \citep[see Figs.~28, 29, 35, in ][]{Werner_Uzdensky-2021}.
While a large statistical study is beyond the scope of this work,
this stochastic variation is shown via the three 2D$yz$ simulations with $B_{gz}=0$ (red diamonds and thin dashed lines) in Fig.~\ref{fig:energyConversion}(a).

In 3D (Fig.~\ref{fig:energyConversion}, magenta circles), $U_{Bxy}$ can be released by reconnection {\it and} DKI.
In addition, flux ropes (3D plasmoids) formed by reconnection may decay via the flux-rope kink instability, releasing still more magnetic energy than in~2D (Fig.~\ref{fig:energyConversion}b). 
Because kinking flux ropes decay, reconnection will not halt as in 2D$xy$, although CS thickening (due to DKI and flux-rope instability) slows both DKI and reconnection; all the 3D simulations in Fig.~\ref{fig:energyConversion}(a) show continued but slow energy conversion beyond~$20\:L/c$ \citep{Werner_Uzdensky-2021}.

In Fig.~\ref{fig:energyConversion}(b) we see that, for~$B_{gz}=0$, all dimensionalities yield comparable~$|\Delta U_{Bxy}|\sim 0.2$--$0.3U_{Bxy0}$.  While a small guide field~$B_{gz}/B_0=1/4$ has only a small effect in~2D$xy$ and~3D, it strongly inhibits energy conversion 
in~2D$yz$ \citep{Zenitani_Hoshino-2008}.

Increasing $B_{gz}$ has two main effects, in all cases: it slows energy conversion, and decreases the net magnetic energy release $|\Delta U_{Bxy}|$.
A guide field of~$B_{gz}/B_0 \gtrsim 1/2$ stabilizes the DKI in~2D$yz$ \citep[cf.][]{Zenitani_Hoshino-2008,Schoeffler_etal-2023}, preventing any release of~$U_{Bxy}$.
For~2D$xy$ and~3D, increasing $B_{gz}/B_0$ from~0 to~1 slows energy conversion and reduces $|\Delta U_{Bxy}|$ by roughly half; by $B_{gz}/B_0=2$, only about~$0.1 U_{Bxy0}$ is released (this depends on the aspect ratio: for~$L_y\gtrsim L_x$, we expect~$|\Delta U_{Bxy}|/U_{Bxy0} \sim 0.1 L_x/L_y$).
Previous 2D$xy$ studies showed that $B_{gz}$ slows reconnection by reducing the Alfv\'{e}n speed $v_{A,x}$ in the outflow direction \citep{Liu_etal-2014,Werner_Uzdensky-2017}; also, $B_{gz}$ 
resists plasmoid compression, halting reconnection at smaller $|\Delta U_{Bxy}|$. 
These trends hold in~3D as well, except that
increasing $B_{gz}/B_0$ from~0 to~$1/4$ releases more energy, possibly because~$B_z$ suppresses the DKI and prevents interference with reconnection, or possibly this is stochastic variation; we leave exploration of this subtrend to future work.

Overall, after $20\:L/c$, $\Delta U_{Bxy}$ is about 15\% higher in 3D than in~2D$xy$ because 3D evolution accesses all the 2D dissipation channels plus the flux-rope instability; the difference between~2D$xy$ and~3D narrows slightly with higher~$B_{gz}$ as guide field increases uniformity in~$z$.  We note that with more time, $|\Delta U_{Bxy}|$ would increase in~3D, but not in~2D$xy$.

\section{Energy gain in guide field} \label{sec:dUBz}

\begin{figure}
\centering
\begin{tabular}{@{}l@{}cl@{}c@{}}
\raisebox{2.1in}{(a)} &
  \includegraphics*[width=3in]{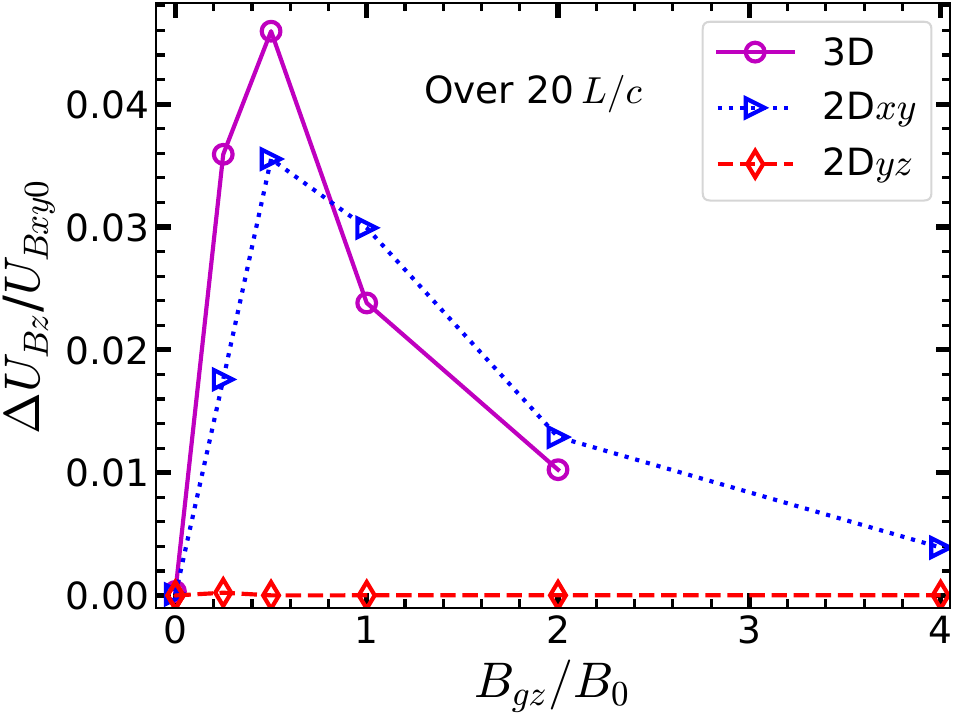}
& \raisebox{2.1in}{(b)} & \includegraphics*[width=3in]{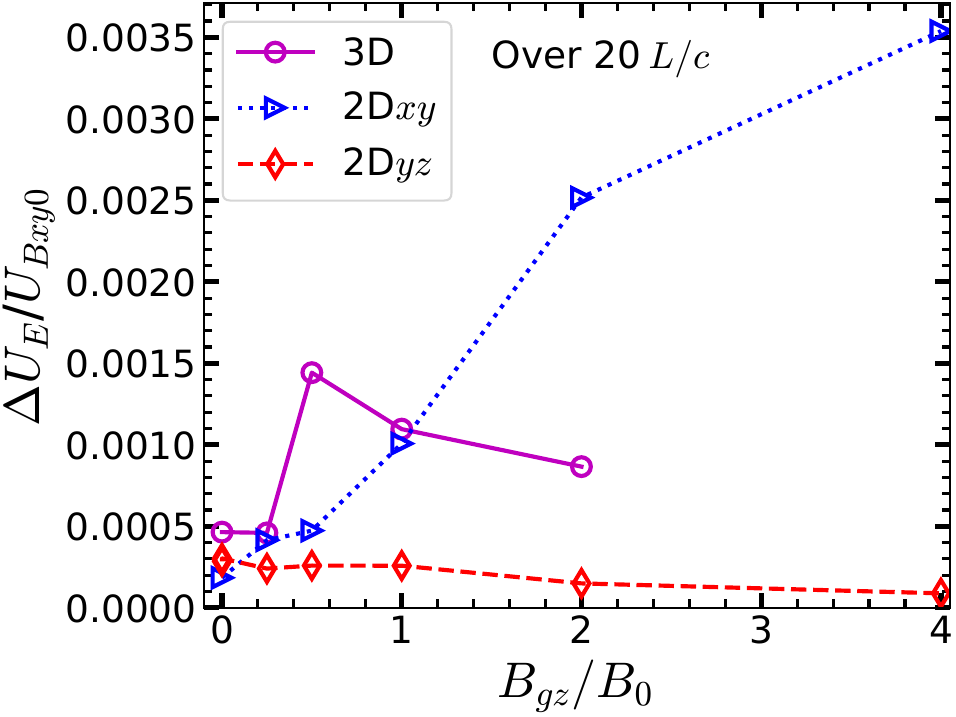}
\\
\raisebox{2.1in}{(c)} & \includegraphics*[width=3in]{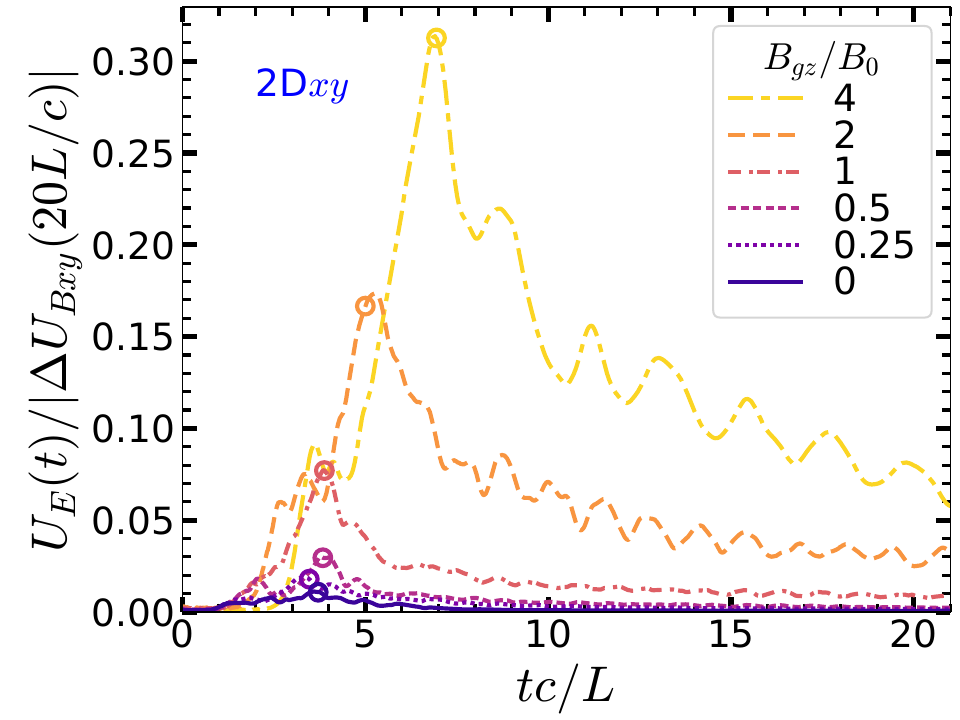}
& \raisebox{2.1in}{(d)}& \includegraphics*[width=3in]{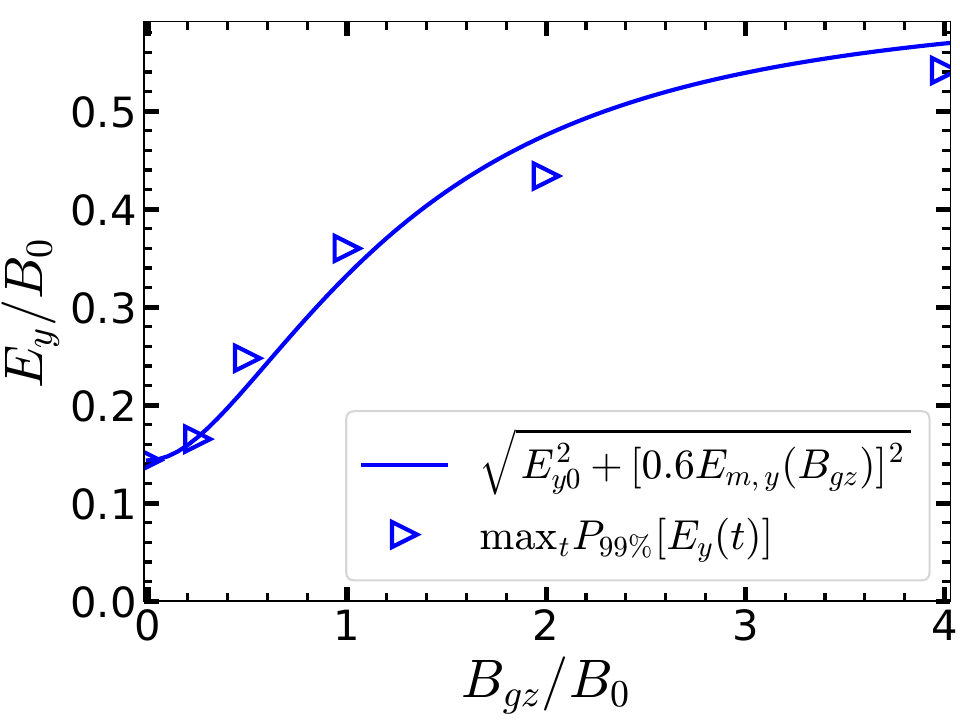}
\end{tabular}
\caption{\label{fig:dUBzdUE}
(a) Net gain in $U_{Bz}$ versus $B_{gz}$, in 2D$xy$ (blue triangles), 3D (magenta circles), and
2D$yz$ (red diamonds showing zero gain).
(b) Net gain in $U_{E}(B_{gz})$.
(c) For 2D$xy$ only, $U_E(t)$ vs. time, for different~$B_{gz}$, normalized to the ultimately-released magnetic energy [not shown: in~2D$yz$ and~3D, $U_E(t)$ remains relatively small].
$U_E$ dissipates only after most of $U_{Bxy}$ has dissipated, as indicated by the circles placed where $\Delta U_{Bxy}(t)=0.75 \Delta U_{Bxy}(20L/c)$.
(d) 
For 2D$xy$, the plasmoid~$E_y$ as a function of~$B_{gz}$, estimated from Eqs.~(\ref{eq:motionalE})--(\ref{eq:Eymax}) (solid line), compared with the maximum $E_y$ measured in simulations (triangles).
(To reduce sensitivity to noise, the
maximum was measured over time of the 99th percentile of~$|E_y|$.)
}
\end{figure}

In our simulations, the guide-field energy $U_{Bz}$ can only increase from its initial value;
the $B_z$-flux through any $x$-$y$ plane is conserved and $B_z$ is initially uniform, yielding the lowest $U_{Bz}$ for the fixed flux.
Figure~\ref{fig:dUBzdUE}(a) shows the net gain $\Delta U_{Bz}$ versus $B_{gz}$ for each dimensionality.
In the 2D$yz$ case, $\Delta U_{Bz} \approx 0$,
and for $B_{gz}=0$ neither 2D$xy$ nor 3D show noticeable gain.
However, for $B_{gz}>0$, reconnection outflows compress plasma {\it and guide field} in flux ropes, increasing~$U_{Bz}$.
The gain $\Delta U_{Bz}$ increases with $B_{gz}/B_0\lesssim 0.5$ because there is more $B_z$ to be compressed; however, stronger $B_{gz}$ rivals the plasma in resisting compression \citep[with adiabatic index 2, greater than the adiabatic index 4/3 or 5/3 of relativistic or nonrelativistic plasma; cf.~\S4.4 of ][]{Werner_Uzdensky-2021}, and~$\Delta U_{Bz}$ peaks around~$0.04\,U_{Bxy0}$ at~$B_{gz}/B_0\sim 0.5$ before decreasing.
The similarity between 2D$xy$ and~3D suggests reconnection occurs to similar extents---though slower in~3D---resulting in similar compression of flux ropes, with energy remaining in $U_{Bz}$ even as flux ropes decay in~3D.

We note that the contribution of the Hall quadrupole $B_z$ field  
is negligible on global energy scales, because it is localized to small regions around X-lines; in contrast, the $U_{Bz}$ in plasmoids becomes globally significant because it grows with plasmoids, which approach the system size.

\section{Energy gain and subsequent loss in electric field} \label{sec:dUE}

The net gain in electric field energy $U_E$ is negligible in all cases---less than $0.004 \,U_{Bxy0}$
(Fig.~\ref{fig:dUBzdUE}b).
For 2D$yz$ and 3D, $U_E(t)$ is small at all times.
However (see Fig.~\ref{fig:dUBzdUE}c), in 2D$xy$ with strong guide field, $U_E(t)$ builds up substantially over time before declining (after most reconnection has occurred); this effect increases with~$B_{gz}$.
For $B_{gz}/B_0=2$, 17\% of the ultimately-released magnetic energy resided \textit{at one time} (around $t=5L/c$) in~$U_E$, and for $B_{gz}/B_0=4$, over~30\%.

This effect---by which substantial $U_{Bxy}$ is first converted to $U_E$ before energizing particles---has not been reported before, to our knowledge.
We find that (when $B_{gz}/B_0\gtrsim 0.5$) $U_E$ is dominated by~$E_y$ in plasmoids; this is simply the motional electric field 
$E_{m,y} \approx v_x B_z/c$ due to $B_z$ in a plasmoid moving with relativistic speed $v_x$ along the layer.  
This effect should appear in any 2D reconnection simulation with guide field and relativistic outflows, even in pair plasma.
After the last major plasmoid merger, as reconnection is ending, only one large but stationary plasmoid remains, and thus the motional electric field ultimately decays.

We can estimate how $E_{m,y} \approx v_x B_z/c$ 
scales with guide field, assuming that $B_z \sim B_{gz}$ and that plasmoid velocity $v_x$ scales with $v_{A,x}$, the Alfv\'{e}n velocity projected along $x$.
Since the typical reconnection electric field is
$E_{\rm rec} \sim 0.1 v_{A,x} B_0/c$ (for fast collisionless reconnection),
$E_{m,y}/E_{\rm rec} \sim 10 B_{gz}/B_0$, which exceeds unity for even small~$B_{gz}$.
In absolute terms,
$v_{A,x} = c[B_0^2/(4\pi h + B_0^2 + B_{gz}^2)]^{1/2}$,
where $h$ is the relativistic plasma enthalpy density including rest-mass 
\citep{Liu_etal-2014,Sironi_etal-2016,Werner_Uzdensky-2017}, and thus
\begin{eqnarray} \label{eq:motionalE}
  E_{m,y}(B_{gz}) &\sim & \sqrt{ \frac{B_{gz}^2 B_0^2 }{4\pi h + B_0^2 + B_{gz}^2} } 
    \rightarrow
    \left\{ \begin{array}{c@{\textrm{ for }}l}
      E_{m,y} \propto B_{gz} & B_{gz}^2 \ll B_0^2 + 4\pi h \\
      E_{m,y} \sim B_{0} & B_{gz}^2 \gtrsim B_0^2 + 4\pi h 
    \end{array} \right.
.\end{eqnarray}
(In this paper, $4\pi h\approx 4\pi n_b m_i c^2 = B_0^2/\sigma_i = 2 B_0^2$.)

This estimate of $E_{m,y}(B_{gz})$ roughly explains the $B_{gz}$-dependence of~$E_y$ in our~2D$xy$ simulations; Fig.~\ref{fig:dUBzdUE}(d) shows the maximum observed $E_y$ compared with 
\begin{eqnarray} \label{eq:Eymax}
  E_{y,\rm max}(B_{gz}) &=& \sqrt{E_{y0}^2 + [\alpha E_{m,y}(B_{gz})]^2}
,\end{eqnarray}
which adds in quadrature the motional field~$E_{m,y}(B_{gz})$ given by~Eq.~(\ref{eq:motionalE}) and~$E_{y0}\equiv E_{y,\rm max}(0)\approx0.14B_0$, an estimate of $E_y$ from other sources (which we measure in the simulation with~$B_{gz}=0$ and assume it remains similar for $B_{gz}>0$), with empirical factor $\alpha \approx 0.6$ accounting for, e.g., $v_x < v_{A,x}$.

Notably, when $U_E$ dissipates at the end of reconnection, it preferentially energizes electrons, as described next.

\section{Energy gain in electrons and ions} \label{sec:dUptcl}

\begin{figure}
\centering
\begin{tabular}{@{}l@{}cl@{}c@{}}
\raisebox{2.1in}{(a)} &
\includegraphics*[width=3in]{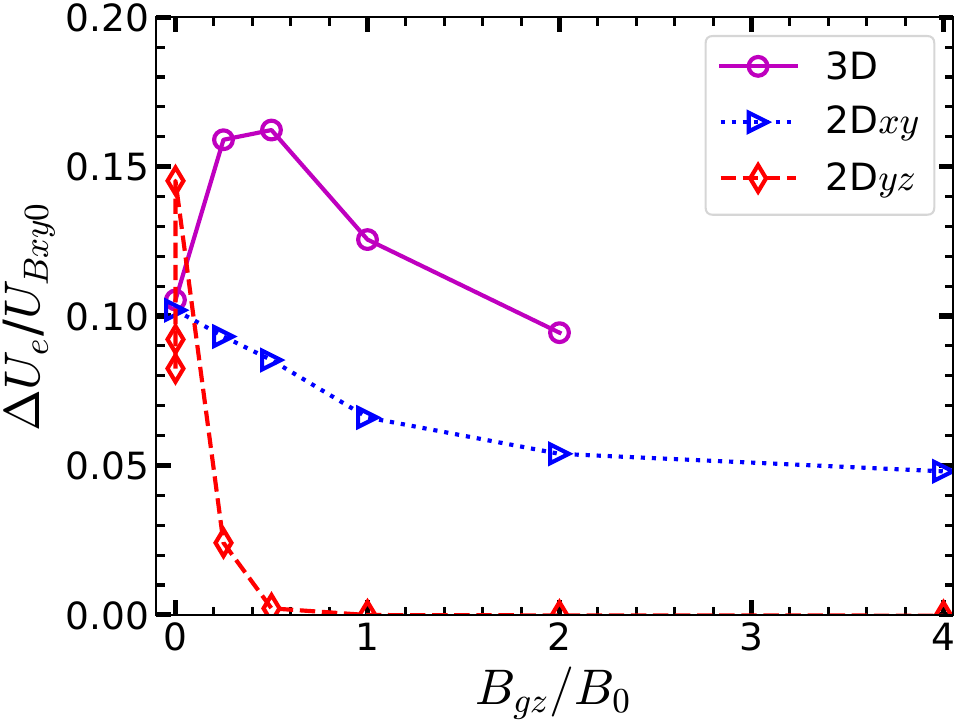}
& \raisebox{2.1in}{(b)} &
\includegraphics*[width=3in]{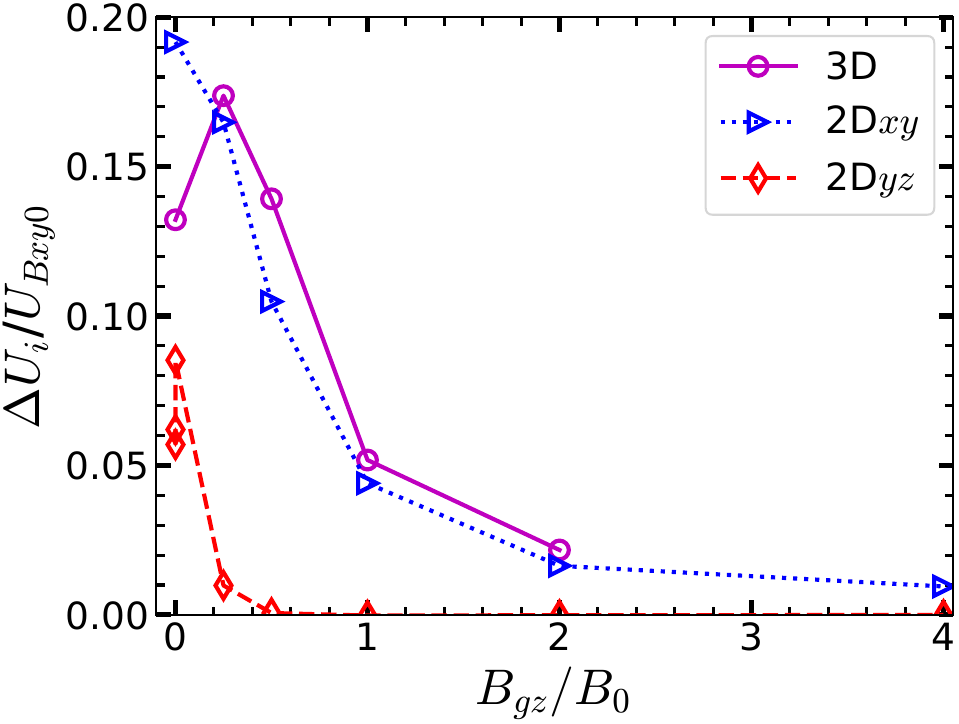}
\\
\raisebox{2.1in}{(c)} &
\includegraphics*[width=3in]{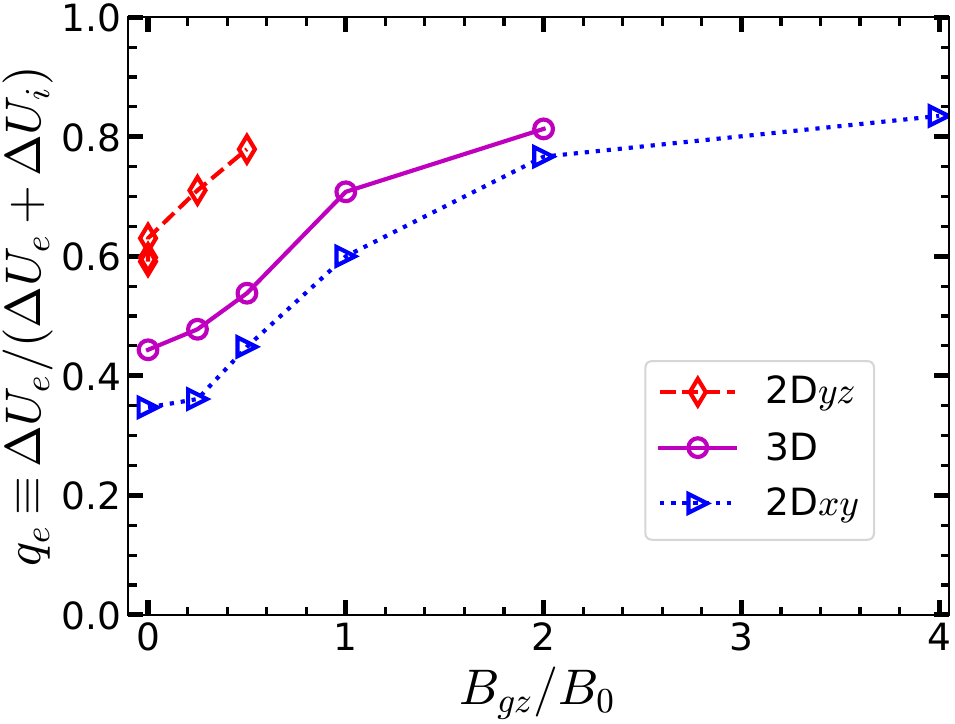}
& \raisebox{2.1in}{(d)} &
\includegraphics*[width=3in]{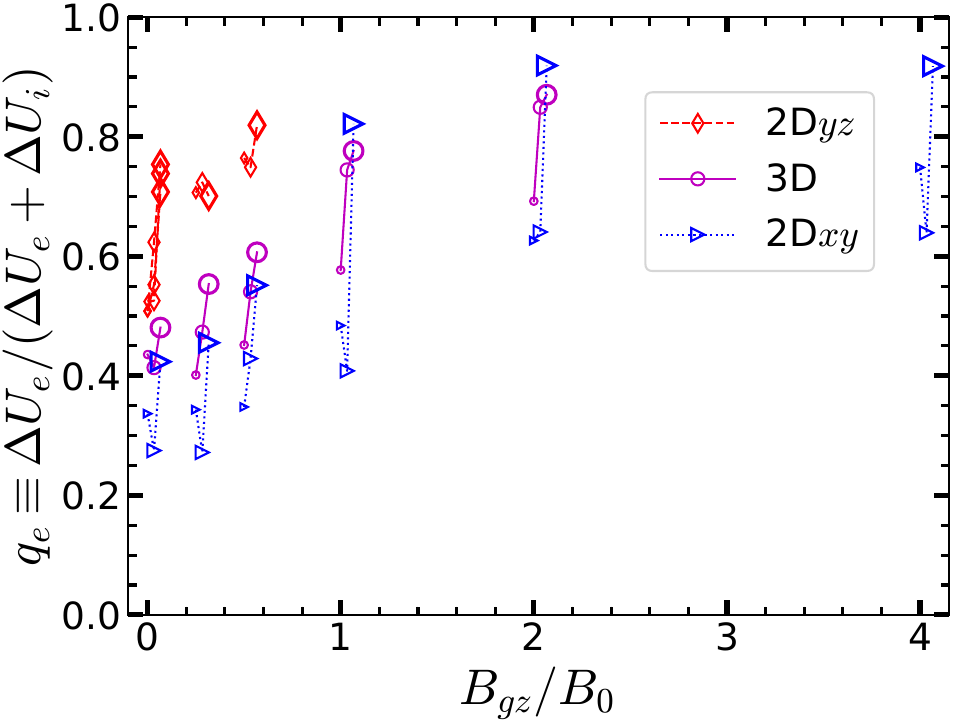}
\end{tabular}
\caption{\label{fig:qe}
The net gain in (a) electron and (b) ion energy, versus $B_{gz}$, normalized by initial transverse magnetic energy~$U_{Bxy0}$,
over the full $T=20\:L/c$.
(c) The electron gain fraction $q_e$ versus $B_{gz}$.
(d) $q_e$ for each third (in terms of $\Delta U_{Bxy}$) of each simulation (the smallest markers, shifted left, show $q_e$ in the first third; the largest, shifted right, show the last third).
2D$yz$ cases with $B_{gz}/B_0>0.5$, hence negligible energization, are omitted.
}   
\end{figure}

In all dimensionalities, increasing guide field generally suppresses both electron and ion energization relative to the available magnetic energy~$U_{Bxy0}$ (Fig.~\ref{fig:qe}a,b).
In~2D$yz$, both electron and ion net gains ($\Delta U_e$ and $\Delta U_i$) drop precipitously with~$B_{gz}$.
In~2D$xy$ and~3D, energy gains also fall, but not to zero, and~$\Delta U_i$ falls much faster than~$\Delta U_e$.
As~$B_{gz}/B_0$ increases from~0.25 to~2, $\Delta U_e$ falls from~$0.09U_{Bxy0}$ to~$0.05U_{Bxy0}$ in~2D$xy$, and from~$0.16U_{Bxy0}$ to~$0.09U_{Bxy0}$ in~3D;
in contrast, $\Delta U_i$ falls from about~$0.17U_{Bxy0}$ to just~$0.02U_{Bxy0}$ (almost an order of magnitude!), and is nearly identical in~2D$xy$ and~3D.
The additional magnetic energy released in~3D (compared with~2D$xy$) goes mostly to electrons.  At low~$B_{gz}$, this may be due to the DKI and/or the flux-rope kink instability; at higher~$B_{gz}$, DKI is quenched, suggesting that the flux-rope kink preferentially energizes electrons. 
The electron energy gain fraction $q_e\equiv \Delta U_{e}/(\Delta U_{e}+\Delta U_{i})$ grows with~$B_{gz}$ (Fig.~\ref{fig:qe}c), even though~$\Delta U_e$ decreases.
In~2D$xy$, $q_e\approx 0.35$ at~$B_{gz}=0$, 
increasing to $q_e\approx 0.5$ (equipartition) around~$B_{gz}/B_0=0.75$, and reaching $q_e>0.8$ by $B_{gz}/B_0=4$.
In~3D, $q_e(B_{gz})$ is higher than in~2D$xy$ by about~0.1 (though this gap narrows as~$B_{gz}$ increases), reaching equipartition near~$B_{gz}/B_0=0.25$, and $q_e\approx 0.7$ at~$B_{gz}/B_0=1$.
In~2D$yz$, $q_e$ is~$\sim 0.2$ higher than in~3D, but we emphasize that, for $B_{gz}/B_0\geq 0.5$, $\Delta U_e$ in 2D$yz$ is tiny compared with~2D$xy$ and~3D (hence hard to measure).

The 3D case with $B_{gz}=0$ is exceptional, converting less energy than higher~$B_{gz}/B_0=0.25$, but the difference is within the stochastic variation observed in 2D$yz$ for $B_{gz}=0$ (cf. Figs.~\ref{fig:energyConversion}b and~\ref{fig:qe}a,b).

In the following, we compare $q_e$ previously measured in~2D$xy$ PIC reconnection simulations in the (nearly) semirelativistic, plasmoid-dominated regime.
For $B_{gz}=0$, $q_e\simeq 0.25$--$0.35$ was found in similar simulations \citep{Rowan_etal-2017,Rowan_etal-2019,Werner_etal-2018,ZhangQ_etal-2021} and also in the different regime of nonrelativistic laboratory experiment \citep[e.g.,][]{Yamada_etal-2014}.
Similar increases in $q_e$ with $B_{gz}$ were also reported in~2D: 
e.g., \citet{Rowan_etal-2019} found $q_e$ increasing to equipartition around $B_{gz}/B_0 \approx 1$ and up to~0.92 for $B_{gz}/B_0=6$;
with reduced mass ratio $\mu=25$,
\citet{Melzani_etal-2014b} found $q_e$ increasing to~0.67 at $B_{gz}/B_0=1$. 
There have, however, been no 3D PIC simulations in a quantitatively comparable regime (to our knowledge) that have measured $q_e$.

Although a detailed analysis of energization mechanisms is beyond the scope of this Letter, the time-dependence of~$q_e$ offers some insight. Figure~\ref{fig:qe}(d) plots $q_e$ separately for each third (in terms of~$\Delta U_{Bxy}$, not time) of each simulation. 
In 2D$yz$, we observe no significant time-dependence of~$q_e$.
For~2D$xy$ and strong~$B_{gz}$, however, the last third (when $U_E$ dissipates; see Fig.~\ref{fig:dUBzdUE}c) has much higher $q_e$ than the first two thirds.  Therefore, we conclude that energization by the motional electric field~$E_{m,y}$ differs from reconnection energization mechanisms that operate earlier (or without~$B_{gz}$), 
and yields higher~$q_e$.
In~3D, for strong~$B_{gz}$, we see the last two-thirds have higher~$q_e$ than the first third, suggesting that $U_E$ builds up a little in~3D, but is promptly dissipated.  That is, energization-via-$E_{m,y}$ operates continuously throughout 3D reconnection, but only at the end of 2D$xy$ reconnection due to the high integrity of plasmoids.
Besides being useful in their own right, measurements of $q_e$ can thus help distinguish between different energization mechanisms.

\section{Summary} \label{sec:summary}

Using PIC simulation, we evolve a thin current sheet in 3D semirelativistic plasma (see Table~\ref{tab:initPlasma}), investigating energy conversion
for different guide magnetic fields~$B_{gz}$.
The 3D simulations are compared with~2D$xy$ (reconnection) and~2D$yz$ 
(drift-kink) simulations with the same initial conditions.

We quantify the net changes in transverse and guide magnetic, electric, electron kinetic, and ion kinetic energies. We find:
\begin{itemize}
  \item Increasing $B_{gz}$ slows energy conversion and reduces the 
  total released magnetic energy. 
  \item Increasing $B_{gz}$ suppresses ion energization more than electron energization, increasing~$q_e\equiv \Delta U_e/(\Delta U_e+\Delta U_i)$.
  In 3D, $\Delta U_{e}/U_{Bxy0}$ falls from~0.16 to~0.09 as $B_{gz}/B_0$ increases from~0.25 to~2, while $\Delta U_{i}/U_{Bxy0}$ falls from~0.17 to~0.02;
  $q_e$ rises from roughly~$0.5$ to~$0.8$.

  \item In 3D only, the flux-rope kink instability further energizes particles (mostly electrons), releasing the reconnected transverse magnetic energy stored in flux ropes, but not their guide-field energy (which is increased by compression during reconnection).

  \item Relativistic reconnection in strong guide field transfers energy to transient motional electric fields~$E_{m,y}$ in flux ropes before preferentially energizing electrons.  In~2D$xy$, $U_E$ grows substantially, and dissipates only at the end of reconnection; in~3D, $U_E$ dissipates promptly, possibly due to flux-rope decay.
  \item Different plasma processes operating in thin CSs yield different electron/ion energy partitions. 
We roughly distinguished electron energy fractions~$q_e$ due to four mechanisms: DKI, classic 2D reconnection, dissipation of~$E_{m,y}$, and flux-rope kink instability. Their relative roles (especially for DKI and reconnection) in~3D may be sensitive to the CS configuration.

\end{itemize}

We have thus characterized the conversion of magnetic energy to electron and ion kinetic energy in~3D CSs, addressing a fundamental plasma-physics question 
and obtaining new insight into the interplay of multiple dissipation channels yielding different energization of electrons versus ions.  Importantly, we measured, from first-principles simulation, electron and ion heating fractions critical for subgrid modeling of electron energization \citep[e.g.,][]{Dexter_etal-2020a,Scepi_etal-2022}, and we showed that they depend significantly on the guide magnetic field.  These results, in combination with similar studies in the 3D relativistic pair-plasma regime \citep[e.g.,][]{Zenitani_Hoshino-2008,Yin_etal-2008,Liu_etal-2011,Kagan_etal-2013,Cerutti_etal-2014b,Guo_etal-2014,Guo_etal-2021,Sironi_Spitkovsky-2014,Werner_Uzdensky-2017,Werner_Uzdensky-2021,ZhangH_etal-2021}, will be important in connecting magnetic energy dissipation with particle energization and observed radiation in the complex and varied enviroment of accreting BHs.

\begin{acknowledgments}

This work was supported by NSF grants AST-1806084 and AST-1903335 and by NASA grants 80NSSC20K0545 and 80NSSC22K0828.
The 3D simulations used computer time provided by the U.S. Department of Energy's (DOE) Innovative and Novel Computational Impact on Theory and Experiment (INCITE) Program, and in particular used resources from the Argonne Leadership Computing Facility, a U.S. DOE Office of Science user facility at Argonne National Laboratory, which is supported by the Office of Science of the U.S. DOE under Contract No. DE-AC02-06CH11357.
The 2D simulations were run on the Frontera supercomputer at the Texas Advanced Computing Center (TACC) at The University of Texas at Austin.
\end{acknowledgments}


\providecommand{\noopsort}[1]{}

\end{document}